\documentclass[11pt]{article}

\usepackage{epsfig} 
\usepackage{graphicx} 

\title{Faster Than Light ?} 

\author{J.\ E.\ Maiorino$^{(a)}$  
\ and \ W.\ A.\ Rodrigues Jr.$^{(b)}$ \\ 
$^{(a)}$Instituto de F\'{\i}sica Gleb Wataghin \\ 
Universidade Estadual de Campinas -
UNICAMP\thanks{maiorino@ime.unicamp.br} \\ 
$^{(b)}$Instituto de Matem\'atica,
Estat\'{\i}stica e Computa\c{c}\~ao Cient\'{\i}fica \\ Universidade
Estadual de Campinas - UNICAMP\thanks{walrod@ime.unicamp.br} }

\date{September 1997} 

\begin{document}

\maketitle 

\begin{abstract} 
In this paper we present a pedestrian review of the theoretical fact that
all relativistic wave equations possess solutions of arbitrary velocities
$0 \leq v < \infty$. We discuss some experimental evidences of $v \geq c$
transmission of electromagnetic field configurations and the importance
of these facts with regard to the principle of relativity.  
\end{abstract} 

\begin{figure} 
\begin{center} 
\input{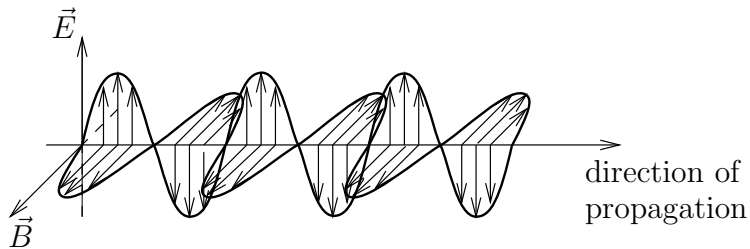}
\end{center} 
\caption{Representation of the oscillating electric 
($\vec{E}$) and magnetic ($\vec{B}$) fields of a plane wave. 
\label{luce1f}} 
\end{figure}

Maxwell's equations are a set of first order partial differential equations
describing the behavior of the electric and magnetic fields generated by
distributions of charges and currents. Such equations also possess solutions for
the case where there are no charges nor currents. The simplest solutions of this
kind, discovered by Maxwell himself, describe electromagnetic field
configurations which we will call \emph{light-solutions} (\emph{LS}).
Light-solutions propagate in empty space with a particular velocity $c$ ($c
\approx 300,000\,\mbox{km/s}$ in MKS units). The simplest, easiest to find
\emph{LS}s are the plane wave solutions (\emph{PW}), which have the
following important characteristics: 

\begin{quote} 
(\emph{i})~\emph{PW}s are \emph{transverse} waves: the electric ($\vec{E}$) and 
magnetic ($\vec{B}$) fields oscillate in time with a certain frequency 
and are perpendicular to the direction of propagation and to each other. 

(\emph{ii}) The field invariants of \emph{PW}s are null. In adequate units, where
$c=1$, these field invariants are given by $I_1 = \vec{E}.\vec{B}$ and $I_2 =
\vec{E}^2 - \vec{B}^2$. 
\end{quote} 
Figure \ref{luce1f} is a pictorial representation of the electric and magnetic
fields of a plane wave. 

One cannot elude a very natural question about \emph{LS}s: What is
the reference frame with respect to which \emph{LS}s propagate with
velocity $c$ ? For Maxwell and his contemporaries the answer  was that
\emph{LS}s have velocity $c$ in a frame ``materialized'' by a special
medium called \emph{aether}, which is their ``carrier'', i.e. \emph{LS}s
are vibrations of this aether. The aether concept was, at their time,
an important part of electromagnetic theory, as indissociable of it as
the concepts of charge and current. The advent of special relativity,
and above all of Einstein's interpretation for it, has taken the aether
concept out of mainstream  physics and into the books on history of
science.\footnote{Contrary to what is written in most textbooks, Einstein
never really abandoned the aether concept, and has even written several
articles in which he claims that this concept is an essential one. See
in this respect the article by L.\ Kostro \cite{kostro}.}

The Earth, our cosmic home, is endowed with several forms of motion
more or less easily detectable. One such motion is its translation
\emph{around} the Sun. It thus cannot be permanently at rest with
respect to the aether and many experiments (of which the most famous is
that of Michelson and Morley) have been carried in order to detect the
effect of the Earth's motion on the measured velocity of light, without
success. These negative results ultimately led to the formulation of
special relativity by Lorentz, Poincar\'e and Einstein.

Before we continue it is important to examine carefully the concept of
velocity itself. Note that in order to determine experimentally the mean
velocity of a particle that traverses a distance $L$ in such a way that
our experiment is a realization of the mathematical definition of this
concept, we must measure the distance $L$ with standard rules and we
shall also need at least two standard clocks. One of them will be at the
beginning of the path $L$ and the other at the end. The clocks must be
synchronized, that is, we must be sure that when the first clock
displays a time, say, $t=0$, the other will register the same time. 

Probably the most noticeable contribution of Einstein to special
relativity was the realization that such a synchronization process
depended on a definition. Einstein then adopted a definition
(called synchronization `\`a l'Einstein') that took into account the
empirical fact that the time for a light signal to go from a point $A$
to a point $B$ in space, with both points fixed in a given inertial
reference frame,\footnote{For a rigorous definition of inertial frame,
see \cite{rodros}.} is independent of the velocity of the source that
emits the signal. He proposed that the clocks at $A$ and $B$ (in any
inertial reference frame) should be synchronized in the following way:
The observer in $A$ sends a light signal to $B$, where the signal is
instantaneously reflected back to $A$. The observer in $A$ determines the
total traversal time $\tau$ for the path $A \rightarrow B \rightarrow A$,
and asks the observer in $B$ to adjust his clock in such a way that it
would show a time $\tau/2$, i.e. one-half the total time for the path
$A \rightarrow B \rightarrow A$, at the moment when the light signal
arrived there. It can be shown that this definition implies that the
measured velocity of light will be always $c$, in any reference frame,
once the standard rules and clocks have a behavior different from that
presupposed by classical theory. And in fact such a distinct behavior
is exactly that found in nature, with a very good approximation. In
particular, it has been found empirically by the American physicists J.\
C.\ Hafelle and T.\ E.\ Keating, in 1968 \cite{hafkea}, that when two
atomic clocks are synchronized at a given point of space in an inertial
refence frame, if one of them makes a trip and comes back to the initial
point after some time, then it will register a time interval smaller than
the time measured by the clock that remained at rest, and the difference
will be exactly what is needed in order that the measurement of the
velocity of any light-solution will be $c$ in every inertial reference
frame.\footnote{In fact, Hafelle and Keating did the experiment on
Earth, which is a non inertial frame. A full account of their results
requires general relativity to be well understood.} If in two inertial
reference frames the spacetime coordinates of events are determined by
standard rules and standard clocks synchronized `\`a l'Einstein', then
the coordinates of an event as determined in both reference frames will
be related by the famous Lorentz transformations (fig.\
\ref{luce2f}\footnote{Adapted from \cite{cihoje}.}).

\begin{figure} 
\begin{center} 
\input{luce2fc.pstex_t}
\end{center}
\vspace{-1cm}
\caption{\label{luce2f}} 
\vspace{1mm}
\begin{minipage}{\textwidth}{\footnotesize 
How is it possible that the velocity of light is always $c$ in every
inertial reference frame? Suppose that $S$ and $s$ are two inertial
laboratories that move with velocity $V$ with respect to each
other. Both laboratories are equipped with standard clocks synchronized
\`a l'Einstein, and the origins of their coordinate systems coincide
for $T = t =0$. In $S$, the equation of motion for a light signal
emitted at $T = t =0$ is written $c^2 T^2 - X^2 -Y^2 - Z^2 = 0$. In
$s$, the equation for the same signal is $c^2 t^2 - x^2 -y^2 -z^2 =0$.
These equations (together with a few other reasonable hypotheses) imply
that the relation between $t$ and $T$, on one hand, and between $x$ and
$X$, on the other, cannot be those used in the Newtonian theory of
spacetime. Indeed, it can be shown that the transformations relating the
coordinates of any event $e$ (e.g. the collision of two particles) in
systems $S$ and $s$, and such that both equations for the light
signals are true, are }
\begin{minipage}{\textwidth}{\footnotesize 
\begin{eqnarray*} 
&&t = \frac{T - VX/c^2}{\sqrt{1- V^2/c^2}}, \quad 
x = \frac{X-VT}{\sqrt{1-V^2/c^2}} ,\quad 
y=Y, \quad z=Z; \\  
&&T = \frac{t + Vx/c^2}{\sqrt{1- V^2/c^2}}, \quad 
X = \frac{x+Vt}{\sqrt{1-V^2/c^2}} ,\quad 
Y=y, \quad Z=z. 
\end{eqnarray*} } 
\end{minipage}  
{\footnotesize Notice that the coordinates of event $e$ are $e= (T, X,
Y, Z)$ in $S$ and $e = (t, x, y, z)$ in $s$. These are the famous Lorentz
transformations, which allow us to calculate, from the measurements made
by a certain observer, the results that would be obtained by another
observer whose state of motion relative to the first one is known, if he
observed the same phenomenon. For $c \rightarrow \infty$ (or for $V/c \ll
1$) these transformations reduce to the Galilean transformations ($x =
X -VT$, $t=T$, $y=Y$, $z=Z$) used in Newtonian theory of spacetime
and in our quotidian calculations, which involve systems endowed with
velocities much smaller than the velocity of light.}
\end{minipage} 
\end{figure}

All these ideas were incorporated by Poincar\'e, in 1904, and Einstein,
in 1905, in what is now known as the theory of special relativity.
Poincar\'e (and also Einstein) assumed the validity of a universal
principle, called principle of relativity, which establishes that the
development of all natural phenomena does not depend on the state of
uniform motion of the inertial reference frame where they take
place.\footnote{The precedence of Poincar\'e over Einstein is well
documented. The interested reader may look at \cite{tai} for details.} 
It can be shown that, from the mathematical point of view, this implies
that all natural phenomena shall be described by equations which posses
the Lorentz group as their symmetry group. Moreover, one can show that
the validity of the principle of relativity implies that no internal
process of synchronization of clocks in a given inertial frame (i.e.
synchronization without ``looking outside the laboratory'') will differ
from the synchronization \`a l'Einstein \cite{rodtio}. 

About his attempt to formulate the principle of relativity, Einstein says
in his \emph{Autobiographical Notes}: ``After ten years of reflection
such a principle resulted from a paradox upon which I had already hit
at the age of sixteen: If I pursue a beam of light with the velocity $c$
(velocity of light in a vacuum), I should observe such a beam of light
as a spatially oscillatory electromagnetic field at rest. However,
there seems to be no such thing, whether on the basis of experience or
according to Maxwell's equations.''

Well, the fact is that Einstein was mistaken. Maxwell's equations are a
source of big surprises. Indeed, just ten years after the publication of
Einstein's fundamental article, the American mathematician H.\ Bateman,
showed in his book Electrical and Optical Motion \cite{bateman} that
the scalar wave equation has solutions that describe a non spreading
packet that travels with speed less than that of any \emph{LS}~! Also,
the late professor A.\ O.\ Barut, of Boulder University, published
in 1992 an extraordinary article \cite{barut} where he showed that
Maxwell's equations without sources also possess wave packet solutions
that travel with velocities less than that of \emph{LS}s. These solutions
present a little dispersion, but it can be shown that in many cases the
time for the spreading of the packet is comparable to the presumed age
of the universe.  Such packets might eventually be used to represent
elementary particles, which would turn out to be nothing but special
electromagnetic configurations. This idea was developed in \cite{rodlu}.

\begin{figure} 
\begin{center}
\includegraphics{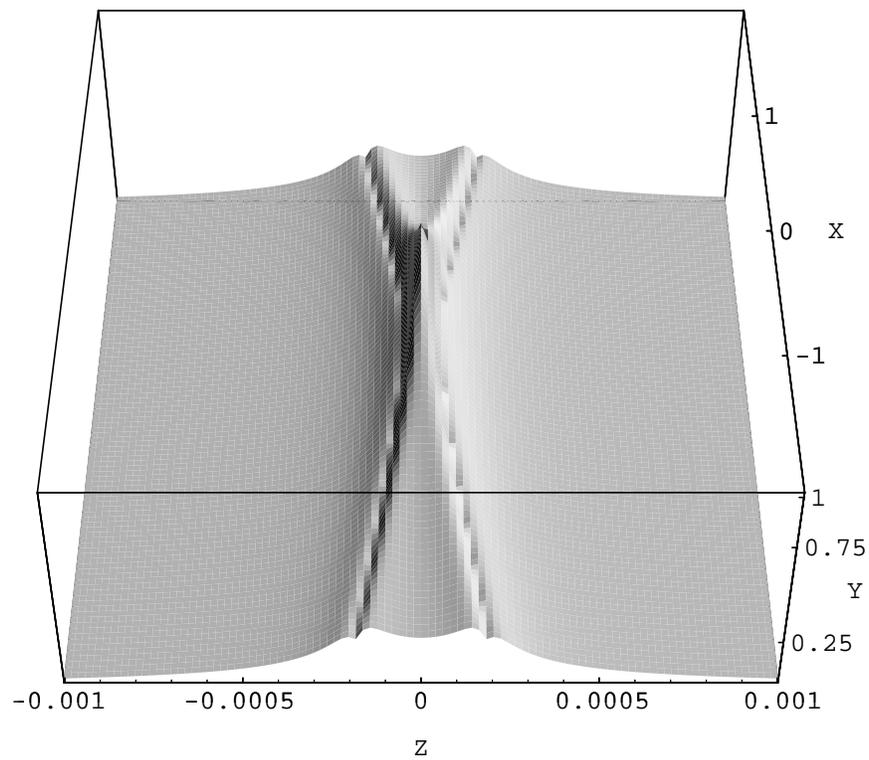}
\end{center} 
\caption{Exact $X$ wave solution of homogeneous wave equation. The
solution is cylindrically symmetric around the $z$ axis; what we
show here is the value of its real part on the plane
$y=0$. \label{luce2af}}
\end{figure}

\begin{figure} 
\begin{center} 
\includegraphics[natheight=5.05954in,natwidth=6in,scale=0.83]{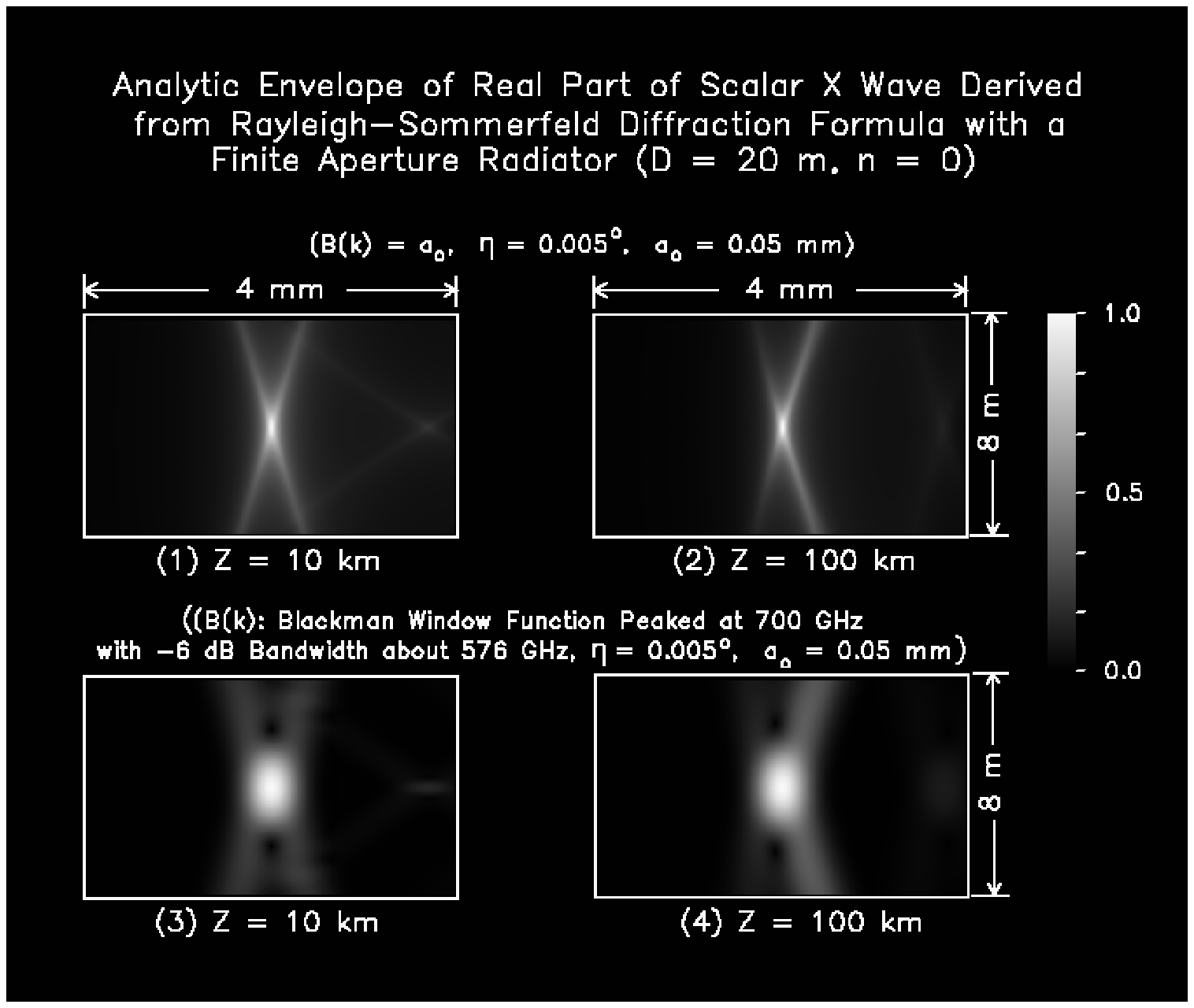}
\end{center}
\caption{Aspect of an$X$ wave propagating in the $z$ direction. The
pictures in (1) and (2) show an approximate solution obtained from the
Rayleigh-Sommerfeld diffraction formula for a broad band $X$ wave; in
(3) and (4) we have the same kind of approximation for a band limited
$X$ wave. The latter might in principle be produced with present day
technology. (Reprinted from \protect\cite{rodlu}.) \label{luce3f}} 
\end{figure}

But do Maxwell equations predict the existence of any electromagnetic
field configuration that propagates with a velocity greater than
$c$? The surprising answer to this question is \emph{yes}. It has been
recently proved \cite{rodvaz1,rodlu,rodmai} that all relativistic wave
equations---the scalar wave equation, Maxwell's equations, and the
Klein-Gordon, Dirac and Weyl equations---have solutions that propagate
with arbitrary velocities $0 \leq v < \infty$. It has been verified that
Maxwell's equations have, besides \emph{LS}s, solutions corresponding to
electromagnetic field configurations that propagate with superluminal
velocities in vacuum.\footnote{See also \cite{donzi}.} One particular
superluminal solution is the so called electromagnetic $X$ wave, which
does not distort as it propagates. Figure \ref{luce2af} shows the real
part of the  $X$ wave solution for the homogeneous wave equation on
the plane $y=0$, propagating in the $z$ direction. A complete solution
of Maxwell equations may be obtained from it by the Hertz potential
method \cite{rodlu}.  In general, electromagnetic configurations that move
(in vacuum) with velocities $v \neq c$ have a longitudinal component
(electric and/or magnetic) and possess at least one non null field
invariant. The exact electromagnetic $X$ wave has infinite energy (as
is the case for plane wave solutions) and thus cannot be produced in
practice. Nevertheless, computer simulations of \emph{finite aperture
approximations} for the $X$ wave, which have finite energies, have shown that these approximations also
propagate with velocity greater than that of \emph{LS}s. These approximate
solutions distort a little as they propagate. Fig.\ \ref{luce3f} shows
the results of such simulations for two different approximations for an
$X$ wave.

\begin{figure}
\begin{center} 
\includegraphics[bb= 113 317 490 513,clip]{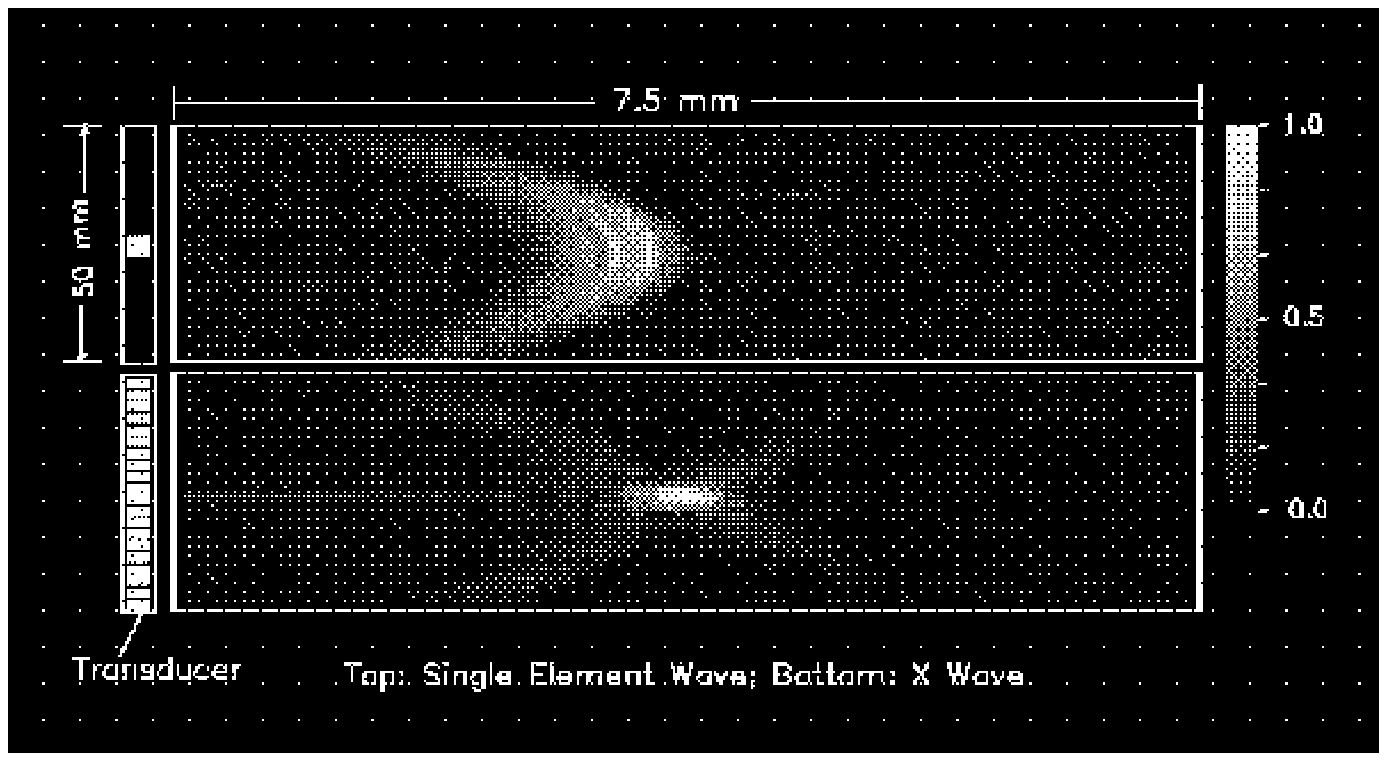}
\end{center} 
\caption{Comparison of the velocities of a single element
acoustic wave and an acoustic $X$ wave. (Reprinted from
\protect\cite{rodlu}.)\label{luce4f}} 
\end{figure}

\begin{figure}
\begin{center}
\includegraphics[bb= 116 320.29 501.14 519.6043,clip]{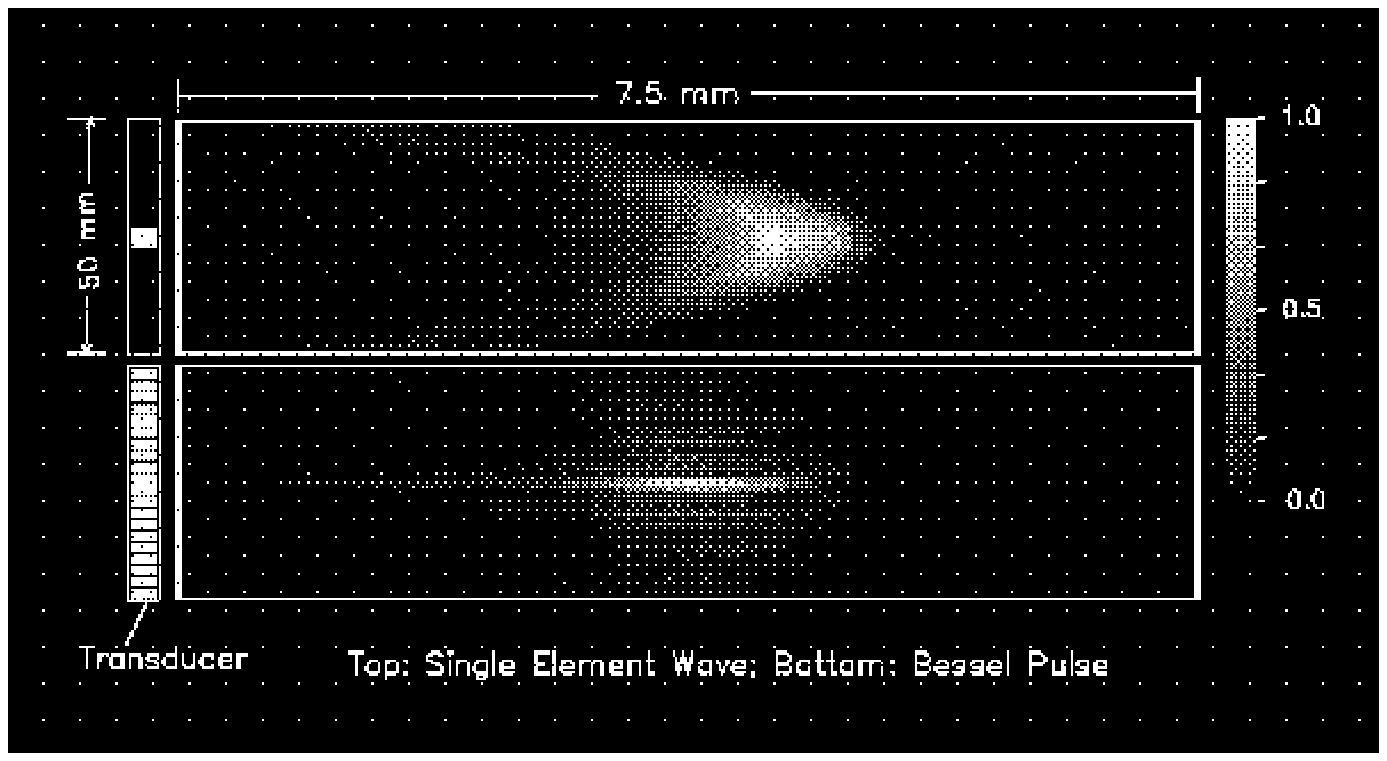}
\end{center}
\caption{Comparison of the velocities of a single element
acoustic wave and an acoustic Bessel pulse. (Reprinted from
\protect\cite{rodlu}.)\label{luce5f}}

\end{figure}

Nobody has so far produced an approximate electromagnetic $X$ wave.
However, there are serious reasons to believe that this is possible. One
of them is the production of (approximate) acoustic $X$ waves, described
in the article by Rodrigues and Lu \cite{rodlu}. Acoustic waves satisfy
a scalar wave equation where the parameter $c$ is replaced by $c_s$,
the velocity of sound. It has indeed been verified that acoustic $X$
waves can travel with speeds greater than $c_s$. Also, another kind of
non spreading acoustic waves called \emph{Bessel pulses} were produced,
which traveled with velocity less than $c_s$, as predicted by the theory
(figs.\ \ref{luce4f} and \ref{luce5f}) .

One may also find theoretical predictions of superluminal propagation
of electromagnetic configurations in several situations which are
mathematically modelled through boundary value problems. As examples
we cite:

\begin{description} 

\item (\emph{i}) Maxwell's equations, when we take into account the
quantum theory of fields. If one solves these equations for the
electromagnetic field inside the region limited by two perfectly
reflecting mirrors with conducting surfaces, one verifies that the
velocity of the electromagnetic field is greater than that of \emph{LS}s
(in vacuum) in the direction perpendicular to the surface of the
mirrors. This solution has been found by G.\ Barton and K.\ Scharnhorst
\cite{barsch}. 

\item (\emph{ii}) One can show that under appropriate boundary conditions
it is possible to generate wave packets that propagate with superluminal
velocities outside a conducting cylinder. This result has been discovered
by W.\ Band \cite{band}. In 1988, P.\ T.\ Papas and A.\ G.\ Obolensky
claimed to have sent signals through coaxial cables with velocities up
to 100 times greater than that of \emph{LS}s \cite{papobo}. Few people
believe in these results, but a generalization of Band's theory predicts
the possibility of propagation of superluminal modes in coaxial cables
under appropriate boundary conditions.

\item (\emph{iii}) Besides these facts, there exist evidences that
microwaves have been launched through horn antennas in the air with
velocities around $1.47\,c$ for distances of about 1~m. For details see
\cite{giaish,ish,ranfabpazmug}.

\end{description} 

Another important theoretical consideration is the following. For more
than half a century the problem of tunneling of wave packets through
potential barriers has been investigated in countless articles. This
problem is an important one since the tunneling of elementary particles of
matter is a nontrivial prediction of quantum mechanics, responsible by
the functioning of several semiconductor devices which are fundamental
for modern technology.\footnote{For a revision see \cite{olkrec}.}
One conclusion of these works is that in the case of the tunneling of
electromagnetic wave packets (which is formally equal to the quantum
mechanical problem) the potential barrier can be physically realized
by a special wave guide where there naturally occur certain modes of
propagation called ``evanescent''.

Several recent experiments have confirmed the superluminal propagation
through barriers. These results have been published in prestigious
periodicals such as \emph{Physical Review Letters}. \emph{Physics
Letters A}, \emph{Journal of Applied Physics} and others. In particular,
R.\ Chiao and his collaborators, from Berkeley University, could
observe a single photon going through a barrier with a velocity 1.47
times the velocity of \emph{LS}s \cite{chiao}. G.\ Nimtz \cite{nimtz}
transmitted Mozart's symphony \#40 between two points 11.7~cm apart
with a velocity 4.7 times that of \emph{LS}s. 

We must remark here that the always quoted results from Sommerfeld and
Brillouin \cite{brisom} showing that electromagnetic waves cannot travel
through a dispersive medium with velocity greater than $c$ is valid only
for \emph{ideal} (or mathematical) signals which have discontinuities
in their first (or second) derivatives and are transverse waves. In
such cases, as is well known (\cite{couhil}, vol.\ 2, p.\ 178), the
signal must propagate along the light-cone characteristics of Maxwell
equations.\footnote{Maxwell equations have more general characteristics
other than the light-cone. See \cite{couhil}.} As correctly pointed out
by Nimtz \cite{nimtz}, ideal signals cannot be produced in practice,
and for \emph{real} signals the Brillouin-Sommerfeld results do not apply.

These spectacular achievements have given rise to much discussion. Are
such results incompatible with the theory of relativity~? Have we found
its limit of validity~? 

Well, we may read in every textbook on special relativity, and also on
research articles, that the theory or relativity implies that no
\emph{signal} may propagate with velocity \emph{greater} than the
velocity of \emph{LS}s in vacuum. This claim is known as
\emph{causality principle} and its acceptance is due mainly to an
argument of Einstein. We shall consider here a more opportune  version
of the argument. 

\begin{figure} 
\begin{center} 
\input{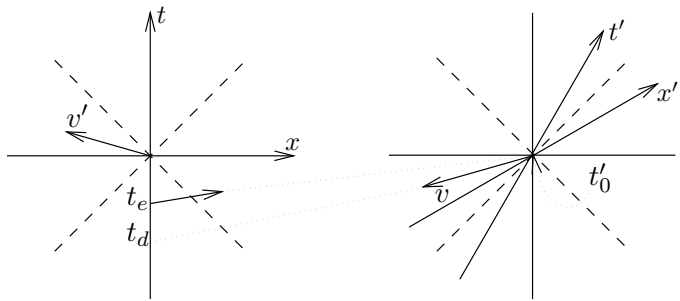}
\end{center}
\caption{Illustration of Einstein's reasoning. The superluminal
signal from $S'$ would arrive at $S$ before the signal from $S$ was
sent. \label{luce6f}} 
\end{figure}

Consider two inertial frames $S$ and $S'$ moving with velocity $V$ with
respect to each other (see fig.\ \ref{luce6f}) and suppose that the observers
in $S$ and $S'$ can produce electromagnetic $X$ waves with velocity $v >
c$, as measured in their respective reference frames, where all clocks
have been synchronized `\`a l'Einstein'. Suppose also that the observer in
$S$ has agreed with his friend in $S'$ to make the following experiment:
``If you receive a signal coming from my laboratory until hour zero of
your clock (time $t_{0'}$ in fig.\ \ref{luce6f}), you should destroy my
laboratory with an $X$ wave of velocity $v>c$, as we have agreed.''  It
can be shown that if $S'$ receives the signal from $S$ (which was sent
at time $t_e$ in fig.\ \ref{luce6f}) and uses his launcher of $X$ waves, he
will be able to destroy the laboratory in $S$ at an instant $t_d$
\emph{earlier than} $t_e$. This constitutes a logical paradox, and based
on such a reasoning Einstein concluded that \emph{there is no
propagation of superluminal signals}. 

Einstein's conclusion does not resist a more careful analysis. Indeed,
in order for this conclusion to be correct it is necessary that relativity
theory itself be valid. But the existence of superluminal signals implies
that there are processes for the synchronization of clocks inside an
inertial reference frame which do not agree with the synchronization \`a
l'Einstein. In particular, the existence of a transcendent superluminal
signal (as is the case for some of the tunneling experiments) may
eventually be used to effect a Newtonian synchronization, i.e. the type
of synchronization supposed true by classical physics, what would imply
a falsification of the principle of relativity \cite{matrod}.

There are many delicate and subtle points about these questions which
cannot be discussed here. We would like to call the readers' attention
to the fact that the claim that there is no contradiction between the
principle of relativity and the existence of superluminal signals, as
maintained by some authors (e.g. \cite{rec}), is false. For a more
thorough discussion of this problem the reader may look at \cite{rodlu}.

Before we finish we would like to remark that even 150 years after their
discovery, Maxwell's equations are still a source of great surprises.
Indeed, in a series of recent articles (\cite{rodvaz1,vazrod,rodvazpav,%
rodvaz2}) it has been verified that there exists an unexpected relation
between Maxwell's equations and the Dirac equation, a relation which
also unveils the geometric origin of the so called supersymmetry between
bosons and fermions. 

We believe that the experiments mentioned above, and a series of other
experiments involving the foundations of quantum mechanics (discussed in
the book \cite{rodgri}), are the first clouds appearing in the horizon of
physics at this end of century. And we believe that they will surely have
unpredictably deeper consequences than those thought of by Lord~Kelvin
at the end of last century, when he referred to the experiments on the
black-body radiation and the Michelson-Morley experiments.


\begin{thebibliography}{99} 

\bibitem{kostro} L. Kostro, in M.\ C.\ Duffy (ed.), \emph{Physical
Interpretations of Relativity Theory}, British Soc. Phil. Sci.  (1989).

\bibitem{rodros} W.\ A.\ Rodrigues Jr. and M.\ A.\ F.\ Rosa,
\emph{Found. Phys.} {\bf 19}, 705 (1989). 

\bibitem{hafkea} J.\ C.\ Hafelle and T.\ E.\ Keating, \emph{Science}
{\bf 177}, 166 (1968); ibid., 168 (1968). 

\bibitem{cihoje} E.\ Recami, M. Fracastoro-Decker and W.\ A.\ Rodrigues
Jr., \emph{Ci\^encia Hoje} {\bf 5}(26), 48 (1986). 

\bibitem{tai} R.\ De La Taille, \emph{Science et Vie} {\bf 951},
114 (1995). 

\bibitem{rodtio} W.\ A.\ Rodrigues Jr. and J.\ Tiomno,
\emph{Found. Phys.} {\bf 15}, 945 (1985). 

\bibitem{bateman} H.\ Bateman, \emph{Electrical and Optical
Motion}, Cambridge, Cambridge Univ. Press (1915). 

\bibitem{barut} A.\ O.\ Barut, \emph{Found. Phys.} {\bf 22},
1267 (1992). 

\bibitem{rodlu} W.\ A.\ Rodrigues Jr.\ and J.-Y.\ Lu, \emph{Found.
Phys.} {\bf 27}, 435 (1997).

\bibitem{rodvaz1} W.\ A.\ Rodrigues Jr.\ and J.\ Vaz Jr.,
\emph{Adv. Appl. Clifford Algebras} {\bf 7} (S), 453 (1997). 

\bibitem{rodmai} W.\ A.\ Rodrigues Jr.\ and J.\ E.\ Maiorino,
\emph{Random Oper. Stoch. Equ.} {\bf 4}, 355 (1996). 

\bibitem{donzi} R.\ Donnelly and R.\ Ziolkowsky, \emph{Proc.
Royal Soc. London A} {\bf 460}, 541 (1993). 

\bibitem{barsch} G.\ Barton and K.\ Scharnhorst, \emph{J. Physics
A} {\bf 26}, 2037 (1993). 

\bibitem{band} W.\ Band, \emph{Found. Phys.} {\bf 18}, 549
(1989). 

\bibitem{papobo} P.\ T.\ Papas and A.\ G.\ Obolensky,
\emph{Electronics and Wireless World} {\bf 12}, 1162 (1988). 

\bibitem{giaish} G.\ C.\ Giakos and T.\ K.\ Ishii, \emph{IEEE Microwave
and Guided Wave Lett.} {\bf 1}, 1051 (1991).

\bibitem{ish} T.\ K.\ Ishii, \emph{Microwaves \& RF}, August 1991,
p.\ 114.

\bibitem{ranfabpazmug} A.\ Ranfagni, P.\ Fabeni, G.\ P.\ Pazzi
and D.\ Mugnai, \emph{Phys. Rev. E} {\bf 48}, 1453 (1993). 

\bibitem{olkrec} V.\ S.\ Olkhovsky and E.\ Recami, \emph{Phys.
Reports} {\bf 214}, 339 (1992). 

\bibitem{chiao} R.\ Chiao et al., \emph{Phys. Rev. Lett.} {\bf
71}, 708 (1993). 

\bibitem{nimtz} G.\ Nimtz, \emph{Phys. Lett. A} {\bf 196}, 154
(1994). 

\bibitem{brisom} L. Brillouin, \emph{Wave Propagation and Group
Velocity}, Academic Press, N.\ Y.\ (1960). 

\bibitem{couhil} R.\ Courant and D.\ Hilbert, \emph{Methods of
Mathematical Physics}, vol.\ 2, Wiley, N.\ Y.\ (1966). 

\bibitem{matrod} T.\ Matolcsi and W.\ A.\ Rodrigues Jr.,
\emph{Algebras, Groups and Geometries} {\bf 15}, 1 (1997). 

\bibitem{rec} E.\ Recami, \emph{N. Cimento} {\bf 9}, 1 (1986). 

\bibitem{vazrod} J.\ Vaz Jr.\ and W.\ A.\ Rodrigues Jr.,
\emph{Int. J. Theor. Phys.} {\bf 32}, 945 (1993). 

\bibitem{rodvazpav} W.\ A.\ Rodrigues Jr., J.\ Vaz Jr.\ and M.\
Pavsic, \emph{Banach Center Publ., Polish Acad. Sciences} {\bf
37}, 295 (1996). 

\bibitem{rodvaz2} W.\ A.\ Rodrigues Jr.\ and J.\ Vaz Jr., in
K.\ Habetha (ed.), \emph{Clifford Algebras and their Applications in Mathematical
Physics}, Kluwer Acad. Publ. (1997). 

\bibitem{rodgri} A.\ A.\ Grib and W.\ A.\ Rodrigues Jr.,
\emph{Nonlocality in Quantum Physics}, to be published. 

\end{thebibliography}
\end{document}